# MICROWAVE AND MILLIMETERWAVE ELECTRICAL PERMITTIVITY OF GRAPHENE MONOLAYER

Alina Cismaru[1], Mircea Dragoman[1*], Adrian Dinescu[1], Daniela Dragoman[2], G. Stavrinidis, G. Konstantinidis[3]

[1] National Institute for Research and Development in Microtechnology (IMT), P.O. Box 38-160, 023573 Bucharest, Romania

[2] Univ. Bucharest, Physics Dept., P.O. Box MG-11, 077125 Bucharest, Romania

[3] Foundation for Research & Technology Hellas (FORTH) P.O. BOX 1527, Vassilika Vouton, Heraklion 711 10, Crete, Hellas

**Abstract**

The effective electrical permittivity of a graphene monolayer is experimentally investigated in the 5-40 GHz range, which encompasses the microwave and the lower part of millimeterwave spectrum. The measurements were carried out using a coupled coplanar waveguide placed over a graphene monolayer flake, which is deposited on $Si/SiO_2$. In contrast to some initial predictions, the effective permittivity of the graphene monolayer is slightly decreasing in the above-mentioned frequency range and has an average value of 3.3.

___________________________________________________________________

*Corresponding author: mircea.dragoman@imt.ro

It is well known that in the optical range graphene has a constant index of refraction of 2.6 in an ultrawide band of wavelengths ranging from ultraviolet up to near-infrared and a constant absorption of 2.3 % (see Ref. 1 and the references therein for a recent review of optical properties of graphene). A similar value of the electrical permittivity of graphene, of almost 3, was estimated from dc measurements of conductivity[2] and changes in the cyclotron mass due to electron-electron interactions,[3] as well as from atomistic calculations for out-of-plane polarization.[4]

However, at microwaves and millimeterwaves, the electrical permittivity plays a key role in designing any device or circuit and its dependence in frequency must be known not from estimations ,but from experimental data. It is not possible to design any graphene device or circuit including field effect transistor amplifiers able to work in microwave and millimeter frequency without the knowledge of effective of electrical permittivity at a certain frequency or in a frequency band, because the the design is based on Maxwell's equations.

The electrical permittivity of graphene in microwave and millimeterwave frequency band is still unknown. Therefore, the role of this paper is to develop an experimental method to find this important parameter in the range 5-40 GHz. The graphene applications in microwave and millimeterwave applications are reviewed in Ref.5.

The information needed to extract the electrical permittivity of graphene in microwaves and millimeterwaves is the dependence on the frequency, $f$, of the phase difference $\Delta\varphi(f)$ between the transmission, $S_{12}(f)$, of a coplanar waveguide (CPW) deposited over graphene and the transmission of an identical CPW in a reference structure, without graphene under the CPW electrodes. Therefore, we have fabricated first a standard graphene CPW configuration, as that illustrated in Fig. 1(a), and its corresponding reference structure. After measuring the phases of both structures with a vector network analyzer (VNA), we have observed that $\Delta\varphi$ is very small, not exceeding 2-3°, and comparable to the phase shifts reported in Ref. 7. To increase $\Delta\varphi$, we decided to choose the configuration in Fig. 1(b), expecting that the phase difference between the coupled CPW reference structure (no graphene) and the coupled CPW with graphene would be larger due to the coupling between the two CPWs. In this structure, the discontinuity of the central electrode exposes a region where the graphene is uncovered by metals and the electric field lines are coupling through it.

As shown in Figs. 2(a) and 2(b), the coupled CPW line consists of continuous outer metallic electrodes (ground electrodes, denoted by G), and a discontinuous central one (signal electrode, denoted by S), which consists of two coupling parts. The coupling distance between the central lines is 1 µm, and the coupling length is 36 µm. The graphene device fabrication process is similar with the graphene device reported recently by us in Ref.6 and will be not repeated here. The fabrication process is reproducible because is a standard clean room process based on e-beam lithography. However, there is a main distinction between graphene device fabrication and standard fabrication of Si devices i.e. before fabrication, Raman spectroscopy of the graphene surface is compulsory to verify: (i) if the graphene is monolayer (the ratio between the intensities of 2D and G peaks is grater than 1.5) and (ii) if the number of defects are low (the absence of D band). A recent review about Raman spectroscopy of graphene is Ref.8. Only after two prerequisites are satisfied a graphene device can be fabricated.

Our device satisfies the above conditions verified by Raman spectroscopy i.e. the ratio of 2D and G peaks were around 2 , while the D band was absent. These data were obtained using a Labram Hr800 Raman spectrometer, having the laser wavelength of 633 nm. The graphene monolayer deposition was provided by Graphene Industries.

We have measured the microwave and millimeterwave response of the graphene device in Fig. 2 using a probe station connected directly with a VNA) Anritsu-37397D. The SOLT (Short, Open, Load, and Thru) calibration standard was used to calibrate the VNA before measurements. The graphene structure is DC biased in the range -2 V to + 2 V (the DC voltage is applied between the central conductor and the ground of the CPW) using the internal bias tee of the VNA coupled to a Keithley 4200 semiconductor characterization system (SCS) able to provide controllable voltage/current sources. To verify if the results are reproducible, we have measured several structures on the same wafer, and their S-parameters were stored in the VNA memory. Each time a new measurement was performed, the stored data and actual data were compared to check the reproducibility.

The phase of the coupled CPW structure on graphene and of the reference coupled CPW structure are depicted in Fig. 3. We have measured the phases at various DC voltages in the case of the coupled CPW structure on graphene, but no phase dependence on voltage was observed. Such relative large phases originate only from the coupled line region of the CPW, where the attenuation is very small: 0.018 dB/μm for 1 μm coupling distance.

Because in any propagating geometry of length $L$, including ours, if only the TEM mode is considered, the phase shift is given by $\Delta\varphi = (\omega/c)\varepsilon^{1/2}L$, and the effective permittivity of graphene can be extracted from:

$$\varepsilon_{eff,gr} = (\varphi^2_{gr,CPW} / \varphi^2_{CPW})\varepsilon_{CPW} \tag{1}$$

where $\varphi_{gr,CPW}$ is the phase shift of the coupled CPW on graphene, $\varphi_{CPW}$ is the phase of the coupled reference structure, while $\varepsilon_{CPW}$ is the electrical permittivity of the Si/SiO$_2$ substrate. Formula (1) holds because the coupled CPW on graphene and the coupled reference CPW fabricated directly on the substrate have the same length and both have a very small attenuation; otherwise the effective permittivities become complex. The electrical permittivity is displayed in Fig. 4. We can see that at the low frequency part of the spectrum we have a relative abrupt decrease of the effective permittivity, while between 10 and 40 GHz the effective permittivity has a slow decrease with frequency, with an average value of 3.3. In addition the decrease of the effective permittivity of graphene with frequency follows the trend observed also in Ref. 7, although the values obtained in Ref. 7 for this parameter are much larger than the current estimations based on different methods and reported in Refs. 1-4. So, the graphene monolayer behaves as a low-permittivity dielectric, almost constant over three decades of frequency within the microwave and millimeter spectrum.

It is interesting to note that the dielectric permittivity of graphene behaves similarly with carbon nanotubes (CNTs ) in the same spectral range [9] despite very different experimental setups. Although the CNT mixture electric permittivity is slightly lower than of electric permittivity of graphene the behavior is the same. Why? CNTs are made from rolled-up graphene sheets. The electromagnetic field wavelengths in microwave and millimeterwave range are much larger than the diameter of CNTs of nm sizes so the electromagnetic propagation is done in a "graphene-like" continuous medium. Therefore, the electric permittivity of graphene and CNTs mixtures must have the same behavior.

In conclusion, the graphene in the microwave and millimeterwave frequencies behaves as a low-permittivity dielectric, similar as in dc and in the optical range. Graphene has an almost constant electrical permittivity over three decades of frequency, in the 10-40 GHz range. It is remarkable that graphene has almost the same permittivity over such a huge frequency range; no other material with similar properties is known. The design of graphene devices and circuits based on the results presented here is now possible. Any commercial simulator used for the simplest graphene device design requires the knowledge of the effective electrical permittivity and its dependence on frequency. Now, we can answer to this requirement which is well known for other materials like Si, GaAs, GaN, $SiO_2$ .

**Acknowledgment**

We thank the European Commission for the financial support via the FP 7 NANO RF (grant agreement 318352) and to the grant CNCS-UEFISCDI, project number PN-II-ID-PCE-2011-3-0071, Romania

**Figure captions**

Fig. 1  (a) Standard CPW configuration and (b) coupled CPW lines.

Fig. 2  (a) The SEM photo of the coupled CPW lines deposited on graphene and (b) a detail of the coupling region.

Fig. 3  The phase of coupled CPW on graphene (black triangles) and of the coupled CPW reference structure (red rhombs).

Fig. 4  The dependence of effective electrical permittivity of graphene on frequency.

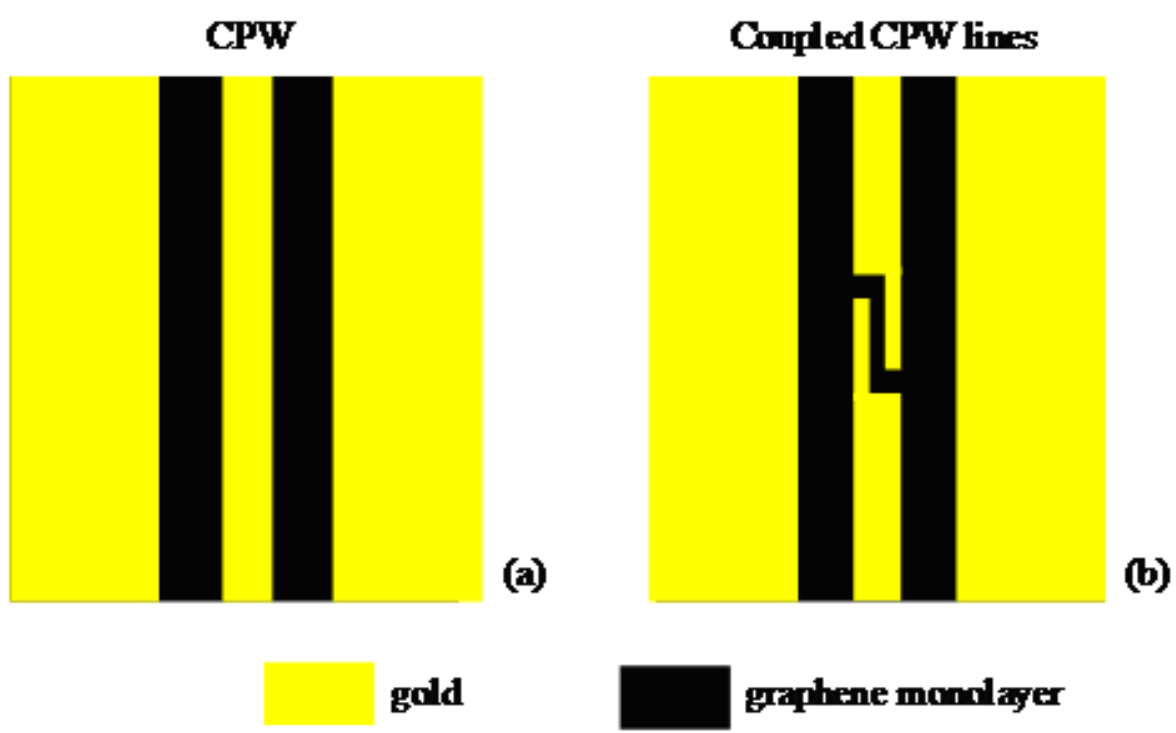

Fig.1

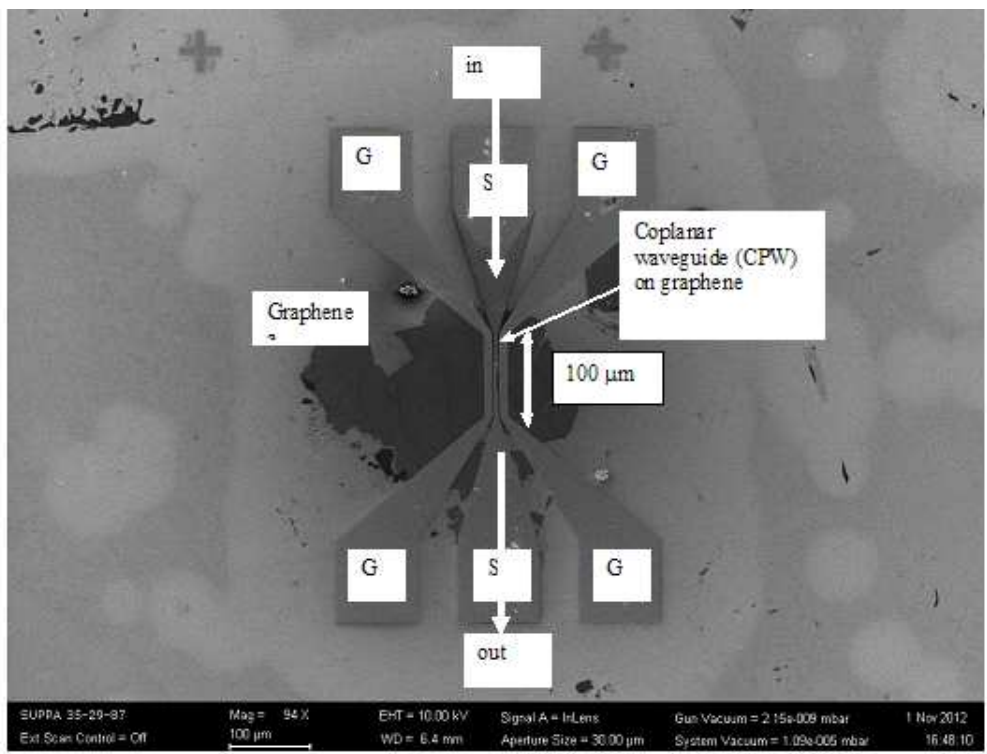

(a)

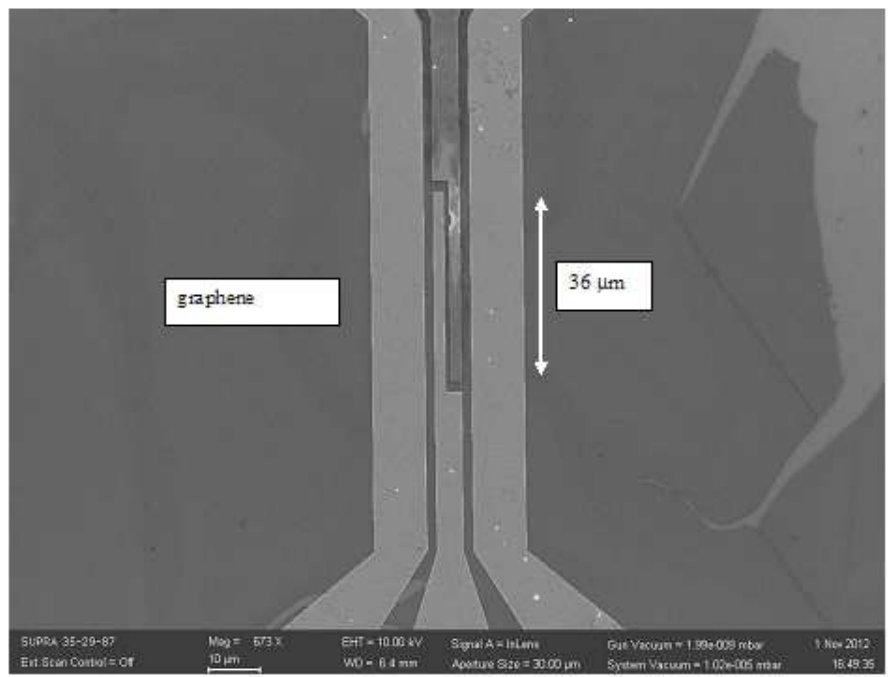

(b)

Fig.2

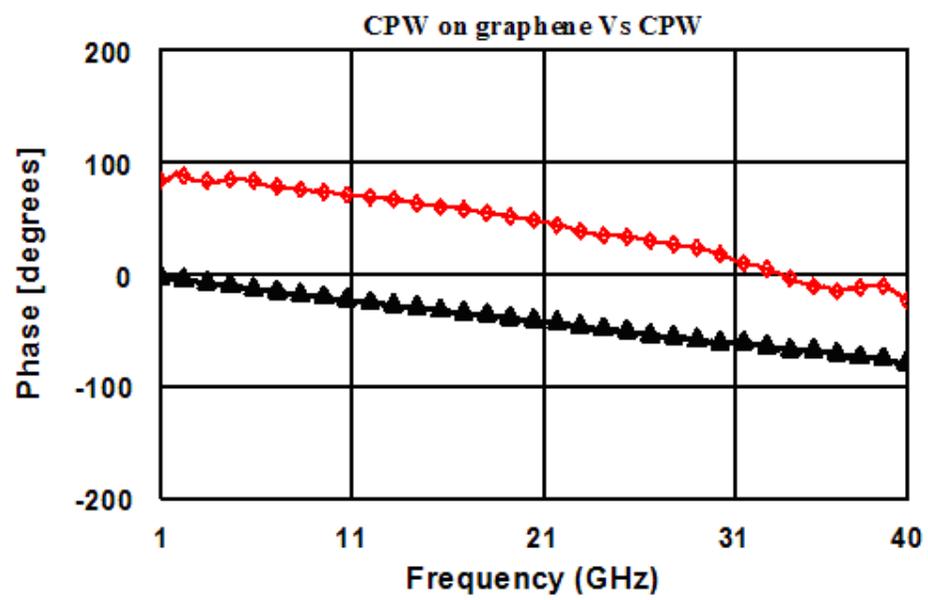

Fig. 3

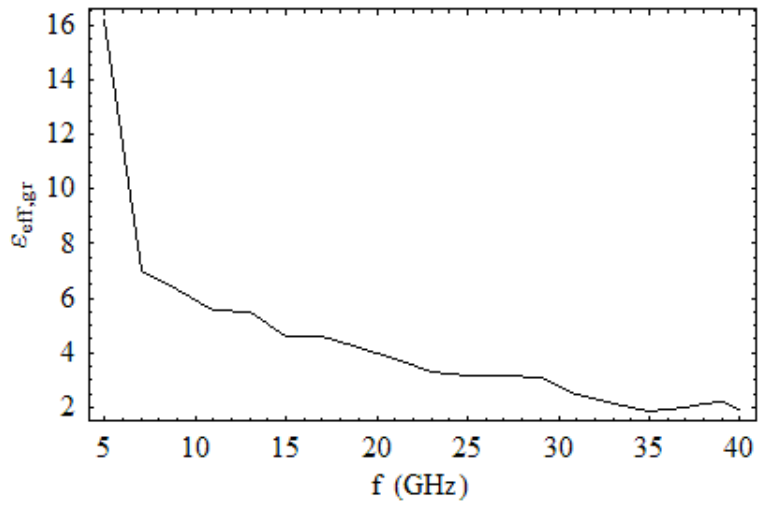

Fig.4